\documentclass[letter,11pt]{article}
\usepackage{jheppub}
\usepackage{tikz}
\usetikzlibrary{decorations.pathmorphing}
\usetikzlibrary{decorations.markings}
\usepackage{xspace}
\usepackage{booktabs}

\usepackage{amsmath}
\usepackage{amssymb}
\usepackage{mathtools}
\usepackage[utf8]{inputenc}
\usepackage{physics}
\usepackage{slashed}
\usepackage{hyperref}
\usepackage[capitalize]{cleveref}

\newcommand{\ep}{\varepsilon}
\newcommand{\g}{\gamma}

\def\3g{{\gamma\gamma\gamma}}
\def\tr5{{\text{tr}_5}}
\newcommand{\tcr}{\texttt{tcr}\xspace}
\newcommand{\tci}{\texttt{tci}\xspace}
\newcommand{\Ffns}{\texttt{F}\xspace}

\title{Two-loop leading-color helicity amplitudes for three-photon production at the LHC}

\author[a]{Herschel A. Chawdhry,}
\author[b]{Micha\l{}  Czakon,}
\author[c]{Alexander Mitov,}
\author[c]{Rene Poncelet}

\affiliation[a]{Rudolf Peierls Centre for Theoretical Physics, Clarendon Laboratory, University of Oxford, Oxford OX1 3PU, United Kingdom}
\affiliation[b]{Institut f\"ur Theoretische Teilchenphysik und Kosmologie, RWTH Aachen University, D-52056 Aachen, Germany}
\affiliation[c]{Cavendish Laboratory, University of Cambridge, Cambridge CB3 0HE, United Kingdom}

\note{Preprint: Cavendish-HEP-20/15, OUTP-20-16P, P3H-20-084, TTK-20-50}

\abstract{We calculate all planar contributions to the two-loop massless helicity amplitudes for the process $q\bar q\to \3g$. The results are presented in fully analytic form in terms of the functional basis proposed recently by Chicherin and Sotnikov. With this publication we provide the two-loop contributions already used by us in the NNLO QCD calculation of the LHC process $pp\to \3g$ [{\it Chawdhry et al. (2019)}]. Our results agree with a recent calculation of the same amplitude [{\it Abreu et al. (2020)}] which was performed using different techniques. We combine several modern computational techniques, notably, analytic solutions for the IBP identities, finite-field reconstruction techniques as well as the recent approach [{\it Chen (2019)}] for efficiently projecting helicity amplitudes. Our framework appears well-suited for the calculation of two-loop multileg amplitudes for which complete sets of master integrals exist.}

\begin{document} 
\maketitle
\flushbottom

\section{Introduction}\label{sec:intro}

Tree-level and multi-loop amplitudes are the building blocks of predictions of gauge field theories and, in particular, for observables in high-energy scattering experiments. For this reason a lot of theoretical effort has been invested in their evaluation and in the development of ever-more efficient methods for their calculation. A number of gauge theories are being studied, starting from string and supersymmetric theories~\cite{Chicherin:2018rpz,Kalin:2018thp,Abreu:2018aqd,Chicherin:2018yne,Abreu:2019rpt,Klemm:2019dbm,Bourjaily:2019gqu,Basso:2020xts,Duhr:2019ywc,Bartels:2020twc,Arkani-Hamed:2019rds,Caron-Huot:2020bkp,Henn:2019swt}, to gravity~\cite{Abreu:2020lyk,Banerjee:2019prz,Chicherin:2019xeg}, to Yang-Mills~\cite{Dalgleish:2020mof,Dunbar:2019fcq}, to QED~\cite{Anastasiou:2020sdt} and QCD~\cite{Larkoski:2017jix,Badger:2018gip,Lim:2018qiw,Abreu:2018zmy,Abreu:2019odu,Hartanto:2019uvl,Ochirov:2019mtf,Dunbar:2020wdh,Poncelet:2020zoc,Czakon:2020qbd,Ahmed:2020nci,Magnea:2020trj,Kardos:2020ppl} and ultimately to the full Standard Model~\cite{Chawdhry:2019bji,Budge:2020oyl,Banerjee:2018lfq,Agarwal:2019rag,Bell:2020qus,Wang:2019mnn,Gehrmann:2020oec,Heinrich:2020ybq,ATLAS:2020xqa,Kallweit:2020gcp}. Research on generic methods for solving multi-loop integrals is also ongoing~\cite{Chawdhry:2018awn,Kotikov:2018wxe,Bosma:2018mtf,Gehrmann:2018yef,Abreu:2018rcw,Boehm:2018fpv,Chicherin:2018mue,Mastrolia:2018uzb,Maierhofer:2018gpa,Kardos:2018uzy,Frellesvig:2019kgj,Bendle:2019csk,Papadopoulos:2019iam,Peraro:2019okx,Guan:2019bcx,Usovitsch:2020jrk,Chicherin:2020oor,Canko:2020ylt,Bendle:2020iim}.

In this work we focus our attention on QCD which is the theory most relevant to high-precision physics at the Large Hadron Collider. Within QCD, the current frontier for $2\to 1$ and $2\to 2$ processes is three loops \cite{Moch:2005tm,Baikov:2009bg,Gehrmann:2010ue,Henn:2013tua,Ahmed:2019qtg,Henn:2020lye,Caola:2020dfu}, while for $2\to3$ scattering processes with massless partons it is two loops. A lot of work has already been carried out in this direction, mostly for planar amplitudes \cite{Abreu:2018gii,Abreu:2018jgq,Badger:2018enw,Abreu:2020xvt,DeLaurentis:2020qle} but recently also for non-planar ones \cite{Chicherin:2018mue,Chicherin:2018old,Badger:2019djh,Guan:2019bcx,Boehm:2020ijp,Klappert:2020nbg,Chicherin:2020oor}. All this progress has enabled the very recent first calculation of a $2\to 3$ process at NNLO, namely, three-photon production at the LHC. This has been achieved by two separate groups \cite{Chawdhry:2019bji,Kallweit:2020gcp} following very different methods. An essential ingredient for these calculations was the corresponding two-loop amplitude $q\bar q\to \3g$. Both references have evaluated it in the leading-color approximation, following different computational approaches. The actual results for the amplitude have, until very recently, not been publicly available.

The goal of the present work is to complete this gap and present the explicit analytic result for the leading-color two-loop amplitude $q\bar q\to \3g$ used in the NNLO calculation of ref.~\cite{Chawdhry:2019bji}. We also compare our result with the one used in ref.~\cite{Kallweit:2020gcp} and published recently in ref.~\cite{Abreu:2020cwb}. We find full agreement between the two results. Our subsequent discussion will be focused on methods used for the evaluation of the amplitude and we refer the interested reader to ref.~\cite{Chawdhry:2019bji} for a broader introduction to the problem and the subject, the definition of the leading color approximation for this process as well as the implications of this calculation. 

The paper is organized as follows. In sec.~\ref{sec:computation} we detail the evaluation of the amplitude. Specifically, in sec.~\ref{sec:notation} we introduce our notation and define the finite remainder; in sec.~\ref{sec:projection} we describe the method for projecting helicities, while in sec.~\ref{sec:reconstruction} we explain how the rational coefficients of the amplitude are derived. Our results are presented in sec.~\ref{sec:results}. They are available for download in electronic form with the arXiv submission of this work.

\section{Computation of the Helicity Amplitudes}\label{sec:computation}

\subsection{Notation and renormalization}\label{sec:notation}

We consider the partonic process
\begin{equation}
  q_c^{h_1}(p_1)\bar{q}^{h_2}_{c'}(p_2) \to \g^{h_3}(p_3)\g^{h_4}(p_4)\g^{h_5}(p_5)\,,
  \label{eq:process}
\end{equation}
where $h_i \in \{+,-\}$ denotes the helicity of the $i$'th parton, $i=1,\dots , 5$. The indices $c,c'$ denote quarks' color. All partons are massless and on-shell
$p_i^2 = 0$.  Momentum conservation and on-shell conditions leave five
independent parity-even Lorentz invariants $s_{ij} = (p_i+p_j)^2$ and one parity-odd 
$\tr5 = 4i\ep_{p_1p_2p_3p_4}$. We choose the following set of
variables to parameterize the amplitudes
\begin{equation}
 x = \{s_{12},s_{23},s_{34},s_{45},s_{51},\tr5\} \;.
\end{equation}
All other Lorentz invariants can be expressed in terms of this set:
\begin{eqnarray}
  s_{13} &= s_{12}-s_{23}-s_{45} \\
  s_{14} &=-s_{15}+s_{23}+s_{45} \\
  s_{24} &= s_{15}-s_{23}+s_{34} \\
  s_{25} &= s_{12}-s_{15}-s_{34} \\
  s_{35} &= s_{12}-s_{34}-s_{45}\;.
\end{eqnarray}
The physical scattering region satisfies \cite{Gehrmann:2018yef}
\begin{equation}
  s_{12} > 0, ~~ s_{12} \geq s_{34}, ~~ s_{45}\leq s_{12} - s_{34}, ~~ s_{23} > s_{12}-s_{45},
  ~~ s_{51}^- \leq s_{51} \leq s_{51}^+ , ~~ (\tr5)^2 < 0\,,
\end{equation}
with
\begin{eqnarray}
 (\tr5)^2 &= &s_{12}^2(s_{23}-s_{51})^2 + (s_{23} s_{34} + s_{45} (s_{34} + s_{51}))^2- \nonumber\\
 & &2 s_{12} (s_{23}^2s_{34} + s_{23} s_{34} s_{45} - s_{23} (s_{34}+s_{45}) s_{51} + s_{45} s_{51} ( s_{34} + s_{51})) \,,
 \label{eq:tr5^2}
\end{eqnarray}
and
\begin{eqnarray}
 s_{51}^{\pm} &=& \frac{1}{(s_{12}-s_{45})^2}\biggl(s_{12}^2s_{23}+s_{12}s_{34}s_{45}-s_{23}s_{34}s_{45} -s_{34} s_{45}^2- \nonumber\\
  & &s_{12}s_{23}(s_{34}+s_{45}) \pm  2 \sqrt{s_{12} s_{23} s_{34} s_{45}(s_{45}+s_{23}-s_{12})(s_{45}+s_{34}-s_{12})} \biggr)\,.
\end{eqnarray}
The UV renormalized amplitude for this
process is denoted by
\begin{equation}
 \mathcal{M}(\alpha_s)_{cc'}^{h_1h_2h_3h_4h_5}(x) =
 \delta_{cc'}
  \mathcal{M}(\alpha_s)^{h_1h_2h_3h_4h_5}(x)
   \equiv \delta_{cc'} \mathcal{M}^{\bar{h}}
\end{equation}
where we factored out the (trivial) color dependence. We summarize the
helicity configuration by $\bar{h} = \{h_1,h_2,h_3,h_4,h_5\}$
and suppress the kinematic dependence for brevity.
The amplitude can be expanded in $\alpha_s$
\begin{equation}
 \mathcal{M}^{\bar{h}} = \mathcal{M}^{\bar{h}(0)}
                    +\biggl(\frac{\alpha_s}{4\pi}\biggr)\mathcal{M}^{\bar{h}(1)}
                    +\biggl(\frac{\alpha_s}{4\pi}\biggr)^2\mathcal{M}^{\bar{h}(2)}
                    +\order{\alpha_s^3}\;.
\end{equation}
The UV renormalized amplitude $\mathcal{M}^{\bar{h}}$ is related to the bare amplitude computed in $d=4-2\ep$ dimensions $\mathcal{M}^{\bar{h},B}$ through
\begin{equation}
 \mathcal{M}^{\bar{h}}(\alpha_s) =
   \left(\frac{\mu^2 e^{\gamma_E}}{4\pi}\right)^{-2\ep}
   Z_q\mathcal{M}^{\bar{h},B}(\alpha_s^0)\,,
\end{equation}
where $Z_q$ is the light quark wave-function renormalization constant.
The bare coupling $\alpha_s^0$ is renormalized in the ${\overline{\rm MS}}$ scheme according to
\begin{equation}
\alpha_s^0 = \left(\frac{e^{\gamma_E}}{4\pi}\right)^{\ep}
             \mu^{2\ep}Z_{\alpha_s} \alpha_s\,.
\end{equation}
All UV renormalization constants are given in appendix \ref{sec:app-renorm}.

The IR divergences of the UV renormalized amplitude can be factorized by means 
of the so-called $\mathbf{Z}$ operator: 
\begin{equation}
\mathcal{M}^{\bar{h}} = \mathbf{Z} \mathcal{F}^{\bar{h}}\,.
\label{eq:IR-fact}
\end{equation}
Once the $\mathbf{Z}$-factor, the finite-remainder
$\mathcal{F}$ and the amplitude $\mathcal{M}$ have been expanded in powers of $\alpha_s/(4\pi)$, eq.~(\ref{eq:IR-fact}) reduces to
\begin{eqnarray}
  \mathcal{M}^{\bar{h}(0)} &=& \mathcal{F}^{\bar{h}(0)}\,,\\
  \mathcal{M}^{\bar{h}(1)} &=& \mathbf{Z}^{(1)}\mathcal{M}^{\bar{h}(0)}
                     +\mathcal{F}^{{h}(1)}\,,\\
  \mathcal{M}^{\bar{h}(2)} &=& \mathbf{Z}^{(2)}\mathcal{M}^{\bar{h}(0)}
                           +\mathbf{Z}^{(1)}\mathcal{F}^{\bar{h}(1)}
                           +\mathcal{F}^{\bar{h}(2)}\,.
\end{eqnarray}
We define $\mathbf{Z}$ in the ${\rm \overline{MS}}$ scheme. 
This completely specifies the finite remainder $\mathcal{F}^{\bar{h}}$.
The explicit expansion for $\mathbf{Z}$ through two-loops in QCD is given in appendix
\ref{sec:app-renorm}. 

The amplitude can be decomposed in color and electric-charge structures. For
the two-loop finite remainder we find four non-vanishing contributions
\begin{eqnarray}
  \mathcal{F}^{\bar{h}(2)}(q\bar q\to \3g) &=& Q^3_q \Bigg(
     4 C_F^2 \mathcal{F}^{\bar{h}, C_F^2}+
     4 C_F C_A \mathcal{F}^{\bar{h}, C_FC_A}+
     2 C_F n_l \mathcal{F}^{\bar{h}, C_Fn_f}
     \Bigg)\nonumber \\
     && +\; Q_q\left(\sum_{q'} Q_{q'}^2\right) 2 C_F \mathcal{F}^{\bar{h}, Q}\,.
     \label{eq:F2-decomposition-Q}
\end{eqnarray}

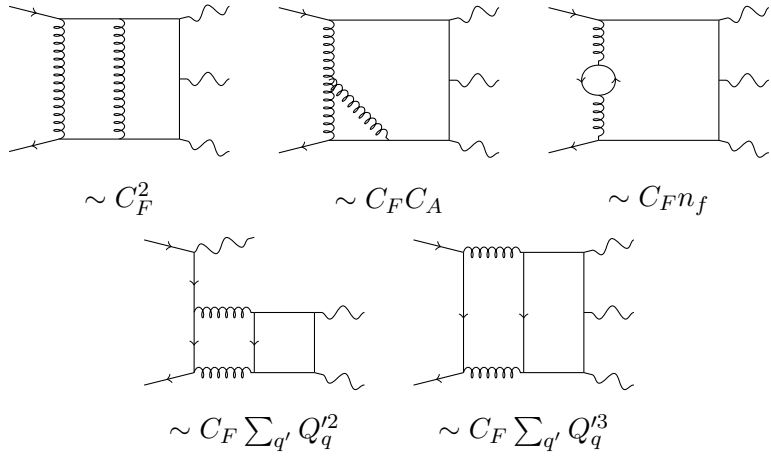
\begin{figure}[]
\centering

\tikzset{
    photon/.style={decorate, decoration={snake}},
    fermion/.style={postaction={decorate},
        decoration={markings,mark=at position .55 with {\arrow[]{>}}}},
    gluon/.style={decorate,
        decoration={coil,amplitude=2pt, segment length=3pt}} 
}

\begin{tikzpicture}[scale=0.4]

\node (caption) at (0,-4) { $\sim C_F^2$};
\node (dummy) at (0,-4.5) { };


%
\node (p1) at  (-4,2.5) {};
\node (p2) at  (-4,-2.5) {};
\node (p3) at  (4,-2.5) {};
\node (p4) at  (4,0) {};
\node (p5) at  (4,2.5) {};

\draw [fermion] (p1) to (-2,2) {};
\draw [fermion] (-2,-2) to (p2) {};
\draw [photon] (p3) to (2,-2) {};
\draw [photon] (p4) to (2,0) {};
\draw [photon] (p5) to (2,2) {};

\draw  (-2,2) to (2,2) {};
\draw  (2,-2) to (-2,-2) {};
\draw  (2,2) to (2,-2) {};

\draw [gluon] (-2,2) to (-2,-2) {};
\draw [gluon] (0,2) to (0,-2) {};
\end{tikzpicture}
\begin{tikzpicture}[scale=0.4]

\node (caption) at (0,-4) { $\sim C_F C_A$};
\node (dummy) at (0,-4.5) { };


%
\node (p1) at  (-4,2.5) {};
\node (p2) at  (-4,-2.5) {};
\node (p3) at  (4,-2.5) {};
\node (p4) at  (4,0) {};
\node (p5) at  (4,2.5) {};

\draw [fermion] (p1) to (-2,2) {};
\draw [fermion] (-2,-2) to (p2) {};
\draw [photon] (p3) to (2,-2) {};
\draw [photon] (p4) to (2,0) {};
\draw [photon] (p5) to (2,2) {};

\draw  (-2,2) to (2,2) {};
\draw  (2,-2) to (-2,-2) {};
\draw  (2,2) to (2,-2) {};

\draw [gluon] (-2,2) to (-2,-2) {};
\draw [gluon] (-2,0) to (0,-2) {};
\end{tikzpicture}
\begin{tikzpicture}[scale=0.4]

\node (caption) at (0,-4) { $\sim C_F n_f$};
\node (dummy) at (0,-4.5) { };


%
\node (p1) at  (-4,2.5) {};
\node (p2) at  (-4,-2.5) {};
\node (p3) at  (4,-2.5) {};
\node (p4) at  (4,0) {};
\node (p5) at  (4,2.5) {};

\draw [fermion] (p1) to (-2,2) {};
\draw [fermion] (-2,-2) to (p2) {};
\draw [photon] (p3) to (2,-2) {};
\draw [photon] (p4) to (2,0) {};
\draw [photon] (p5) to (2,2) {};

\draw  (-2,2) to (2,2) {};
\draw  (2,-2) to (-2,-2) {};
\draw  (2,2) to (2,-2) {};

\draw [gluon] (-2,2) to (-2,0.5) {};
\draw [gluon] (-2,-0.5) to (-2,-2) {};
\draw [fermion] (-2,0.5) to[out=180,in=180,distance=0.75cm] (-2,-0.5) {};
\draw [fermion] (-2,-0.5) to[out=0,in=0,distance=0.75cm] (-2,0.5) {};
\end{tikzpicture}

\begin{tikzpicture}[scale=0.4]

\node (caption) at (0,-4) { $\sim C_F \sum_{q'} Q_q'^2$};
\node (dummy) at (0,-4.5) { };


%
\node (p1) at  (-4,2.5) {};
\node (p2) at  (-4,-2.5) {};
\node (p3) at  (4,-2.5) {};
\node (p4) at  (4,0) {};
\node (p5) at  (4,2.5) {};

\draw [fermion] (p1) to (-2,2) {};
\draw [fermion] (-2,-2) to (p2) {};
\draw [photon] (p3) to (2,-2) {};
\draw [photon] (p4) to (2,0) {};
\draw [photon] (-2,2) to (0,2.5) {};

\draw [fermion] (-2,2) to (-2,0) {};
\draw [fermion] (-2,0) to (-2,-2) {};
\draw [gluon] (-2,0) to (0,0) {};
\draw [gluon] (-2,-2) to (0,-2) {};

\draw [fermion] (0,0) to (0,-2) {};
\draw  (0,0) to (2,0) {};
\draw  (2,0) to (2,-2) {};
\draw  (2,-2) to (0,-2) {};
\end{tikzpicture}
\begin{tikzpicture}[scale=0.4]

\node (caption) at (0,-4) { $\sim C_F \sum_{q'} Q_q'^3$};
\node (dummy) at (0,-4.5) { };


%
\node (p1) at  (-4,2.5) {};
\node (p2) at  (-4,-2.5) {};
\node (p3) at  (4,-2.5) {};
\node (p4) at  (4,0) {};
\node (p5) at  (4,2.5) {};

\draw [fermion] (p1) to (-2,2) {};
\draw [fermion] (-2,-2) to (p2) {};
\draw [photon] (p3) to (2,-2) {};
\draw [photon] (p4) to (2,0) {};
\draw [photon] (p5) to (2,2) {};

\draw [fermion] (-2,2) to (-2,-2) {};
\draw [gluon] (-2,2) to (0,2) {};
\draw [gluon] (-2,-2) to (0,-2) {};

\draw [fermion] (0,2) to (0,-2) {};
\draw  (0,2) to (2,2) {};
\draw  (2,2) to (2,-2) {};
\draw  (2,-2) to (0,-2) {};
\end{tikzpicture}
\caption{Representative two-loop diagrams and their color/charge factors.
}\label{fig-diagrams}
\end{figure}

In the above equation, $Q_i$ are quarks' QED couplings ($Q_{u,c,t} = \frac{2}{3} e$ and 
$Q_{d,s,b} = -\frac{1}{3}e$). The sum over $q'$ in eq.~(\ref{eq:F2-decomposition-Q}) goes over 
all massless quarks. The difference between the QED couplings in the first and second lines of eq.~(\ref{eq:F2-decomposition-Q}) 
is due to the following: all diagrams 
contributing to the first line of eq.~(\ref{eq:F2-decomposition-Q}) have all three photons coupling 
to the external quark line (of flavor $q$) while the diagrams contributing to the second line of eq.~(\ref{eq:F2-decomposition-Q}) 
have one photon coupling directly to the external quark line (of flavor $q$) and 
two photons coupling to an internal fermion loop with a flavor $q'$. 
Example diagrams for the different contributions can be found in fig.~\ref{fig-diagrams}.
In this work we do not consider diagrams with massive quark loops. We also note that the
contribution proportional to $C_F\sum_{q'} Q_{q'}^3$ from diagrams with 
three photons coupling to an internal fermion loop (see fig.~\ref{fig-diagrams})
vanishes by Furry's theorem.

The leading color contribution of $\mathcal{F}^{\bar{h}(2)}(q\bar q\to \3g)$  is proportional to 
$N_c^2$ and, as follows from eq.~(\ref{eq:F2-decomposition-Q}), 
is only dependent on a linear combination of the first two factors in the first line of eq.~(\ref{eq:F2-decomposition-Q}): 
\begin{equation}
  \mathcal{F}^{\bar{h}(2)}(q\bar q\to \3g)\bigg\vert_{\rm l.c.} = Q^3_q N_c^2 \Bigg(
  \mathcal{F}^{\bar{h}, C_F^2}+2\mathcal{F}^{\bar{h}, C_FC_A}\Bigg) + {\cal O}(N_c)\,.
  \label{eq:F2-decomposition-LC}
\end{equation}
The phenomenological analysis in ref.~\cite{Chawdhry:2019bji} is based on eq.~(\ref{eq:F2-decomposition-LC}). 
The justification for the use of this approximation can be found in that reference. 

Despite recent progress, the non-planar diagrams in this process are still beyond reach. 
For this reason, in this work we derive and present in analytic form the planar results for the following two factors: 
\begin{equation}
\mathcal{F}^{\bar{h}, C_F^2}+2\mathcal{F}^{\bar{h}, C_FC_A}~~ {\rm and} ~~ \mathcal{F}^{\bar{h}, C_Fn_f}\,.
\label{eq:color-structures}
\end{equation}

\subsection{Helicity projections}\label{sec:projection}

The process (\ref{eq:process}) has two independent helicity amplitudes. As such we choose
\begin{equation}
  \mathcal{F}^{\{+----\}} \equiv \mathcal{F}^{\bar{h}_-}\quad \text{and} \quad \mathcal{F}^{\{+---+\}}\equiv \mathcal{F}^{\bar{h}_+}\,.
  \label{eq: helicities}
\end{equation}
All other helicities can be obtained from this set by conjugation and/or permutation of external momenta. 
Since at tree-level $\mathcal{F}^{(0),\bar{h}_-} = 0$, the one-loop contribution to this helicity is finite and its two-loop correction does not contribute to the squared matrix element for this process through two loops. 

In order to extract the helicity amplitudes (\ref{eq: helicities}) in the 't Hooft-Veltman scheme we employ the projection method proposed in ref.~\cite{Chen:2019wyb}. Similar approaches have been advocated in refs.~\cite{Peraro:2019cjj,Peraro:2020sfm}. The essence of this method is that, for a given helicity, it provides explicit prescription for constructing the external wave-functions. For the process under consideration, these wave-functions can be factored out by writing the amplitude in the following way
\begin{equation}
  \mathcal{M}^{\bar{h}} = \ep_{3,h_3}^{*\mu}\ep_{4,h_4}^{*\nu}\ep_{5,h_5}^{*\rho}
                 \bar{v}(h_2) \Gamma_{\mu\nu\rho}u(h_1)\,.
 \label{eq:amplitude}                 
\end{equation}

The construction of the external wave-functions is done in $4$ dimensions. 
On one hand this significantly simplifies the construction of a basis of polarization vectors.  
On the other, it introduces scheme dependence into the bare loop amplitude. As
demonstrated in refs.~\cite{Chen:2019wyb,Ahmed:2019udm,Ahmed:2020kme}, however, 
this scheme dependence does not affect the finite remainder of the amplitude in the limit $\ep \to 0$. 

Our construction of the fermionic wave functions introduces the matrix $\gamma_5$. 
The fact that wave functions are constructed in 4 dimensions implies that we only use 4-dimensional identities for $\gamma_5$. With the help of these identities we eliminate all occurrences of $\gamma_5$ and, in this process, trade them for objects involving $\ep^{\mu\nu\rho\sigma}$. Multiplying (and dividing) the term proportional to $\ep^{\mu\nu\rho\sigma}$ by $(\tr5)^2$ and recalling that $(\tr5)^2$ is a polynomial function of the invariants $s_{ij}$ (see eq.~(\ref{eq:tr5^2})), we eliminate all occurrences of $\ep^{\mu\nu\rho\sigma}$ using
\begin{equation}
\begin{split}
 \ep_{p_1p_2p_3p_4} \ep^{\mu\nu\rho\sigma} &= 
+p_1^\rho p_2^\nu p_3^\mu p_4^\sigma
-p_1^\nu p_2^\rho p_3^\mu p_4^\sigma
-p_1^\rho p_2^\mu p_3^\nu p_4^\sigma
+p_1^\mu p_2^\rho p_3^\nu p_4^\sigma
+p_1^\nu p_2^\mu p_3^\rho p_4^\sigma
-p_1^\mu p_2^\nu p_3^\rho p_4^\sigma\\
&-p_1^\rho p_2^\nu p_3^\sigma p_4^\mu
+p_1^\nu p_2^\rho p_3^\sigma p_4^\mu
+p_1^\rho p_2^\sigma p_3^\nu p_4^\mu
-p_1^\sigma p_2^\rho p_3^\nu p_4^\mu
-p_1^\nu p_2^\sigma p_3^\rho p_4^\mu
+p_1^\sigma p_2^\nu p_3^\rho p_4^\mu\\
&+p_1^\rho p_2^\mu p_3^\sigma p_4^\nu
-p_1^\mu p_2^\rho p_3^\sigma p_4^\nu
-p_1^\rho p_2^\sigma p_3^\mu p_4^\nu
+p_1^\sigma p_2^\rho p_3^\mu p_4^\nu
+p_1^\mu p_2^\sigma p_3^\rho p_4^\nu
-p_1^\sigma p_2^\mu p_3^\rho p_4^\nu\\
&-p_1^\nu p_2^\mu p_3^\sigma p_4^\rho
+p_1^\mu p_2^\nu p_3^\sigma p_4^\rho
+p_1^\nu p_2^\sigma p_3^\mu p_4^\rho
-p_1^\sigma p_2^\nu p_3^\mu p_4^\rho
-p_1^\mu p_2^\sigma p_3^\nu p_4^\rho
+p_1^\sigma p_2^\mu p_3^\nu p_4^\rho\,,
\end{split}
\end{equation}
and then promoting all indices to $d$ dimensions. Once this has been achieved, the subsequent contractions of external polarization states with the rest of the amplitude are performed in $d$ dimensions. The remaining factor $\ep_{p_1p_2p_3p_4}$ is converted into $\tr5$ which is then treated as an independent kinematic variable that is included in the rational coefficients $A_b$ defined in sec.~\ref{sec:reconstruction} below.

We next derive explicit expressions for the wave functions of the final state vector particles of given helicity $h=\pm 1$
\begin{equation}
  \ep_{i,h}^{\mu} \quad \text{with} \quad i \in \{3,4,5\}\,.
\end{equation}
As a first step we replace those with new linear polarization vectors
\begin{equation}
  \ep_{i,h}^\mu = \frac{1}{\sqrt{2}}( \ep_{i,X}^\mu +h i \ep_{i,Y}^\mu ) \,,
  \label{eq:wf-vector}
\end{equation}
polarized along two directions $X$ and $Y$. 
The polarization vector along $X$ is defined through the following ansatz
\begin{equation}
  \ep_{i,X}^\mu = c^X_{i,1} p_{1}^{\mu} + c^X_{i,2} p_{2}^{\mu}
              + c^X_{i,3} p_{i}^{\mu}\,.
\end{equation}

As a reference vector for all vectors $\ep_{i,X}^\mu$ we chose the vector $q^{\mu} = p_{1}^{\mu} + p_{2}^{\mu}$. The coefficients $c^X_{i,n}, n=1,2,3$ are determined from the system of normalization and orthogonality conditions for the polarization vectors:
\begin{equation}
  (\ep_{i,X})^2 = -1 \quad,\quad
  \ep_{i,X}\cdot q = 0 \quad,\quad
  \ep_{i,X}\cdot p_i = 0 \;.
\end{equation}

The polarization vector along the direction $Y$ is given by
\begin{equation}
  \ep_{i,Y}^{\mu} = \mathcal{N}_{i,Y}
                    \ep^\mu_{~\nu\rho\sigma}q^{\nu}p_{i}^\rho\ep^\sigma_{i,X}\,,
\end{equation}
where we have used the conventions of ref.~\cite{Chen:2019wyb} for the Levi-Civita symbol: $\ep^{0123} = +1$ and $\ep_{0123} = -\ep^{0123}$. The normalization factors $\mathcal{N}_{i,Y}$ are determined from the condition $\ep_{i,Y}^2 = -1$. 

Lastly, we note that the above construction does not fix the vectors $\ep_{i,X}$ and $\ep_{i,Y}$ uniquely. There are two possible solutions which correspond to the change $\ep_{i,X/Y} \to -\ep_{i,X/Y}$. The overall signs of these two vectors are chosen in such a way that the vectors $\vec{p}_i, \vec{\ep}_{i,X}$ and $\vec{\ep}_{i,Y}$ form a right-handed coordinate system.

The fermion wave functions are treated in the following way. The spinor part of the amplitude (\ref{eq:amplitude}) has the following structure
\begin{equation}
  \mathcal{M} = \bar{v}(p_2,h_2) \Gamma u(p_1,h_1)
              = \Tr \biggl\{ \left(u \otimes \bar{v}\right)
                  \Gamma \biggr\}\,.
\end{equation}

The matrix $u\otimes\bar{v}$ in the above equation can be rewritten in the following way:
\begin{equation}
  (u\otimes\bar{v})_{\alpha\beta} = \frac{\bar{u}Nv}{\bar{u}Nv} (u\otimes\bar{v})_{\alpha\beta}
           = \frac{1}{\bar{u}Nv} (u\otimes\bar{u})_{\alpha\gamma} N_{\gamma\delta}
              (v \otimes \bar{v})_{\delta\beta}
           = \frac{1}{\mathcal{N}}[(u\otimes\bar{u}) N
              (v \otimes \bar{v})]_{\alpha\beta}\,,
  \label{eq:u-times-vbar}
\end{equation}
for some matrix $N$, to be specified below, and $\mathcal{N} \equiv \bar{u}Nv \neq 0$. The outer spinor products read
\begin{eqnarray}
 u(p,h_1) \otimes \bar{u}(p,h_1) = \slashed{p}\frac{1-h_1\g_5}{2}\,,\\
 v(p,h_2) \otimes \bar{v}(p,h_2) = \frac{1-h_2\g_5}{2} \slashed{p}\,.
\end{eqnarray}

The matrix $N$ depends on the process-specific kinematics. In particular, it
is linearly independent of $p_1$ and $p_2$, otherwise $\bar{u}Nv = 0$. 
Since for helicity configurations with $h_1 = h_2$ the amplitude (\ref{eq:process}) 
vanishes to all orders, in the following we only consider the case $h = h_1 = -h_2$.  
For this helicity configuration the matrix $N$ is given by
\begin{eqnarray}
  N &=& i \ep_{\g p_3 p_4 p_5} \quad \text{if} \quad h_1 \neq h_2 \,.
\end{eqnarray}

For this choice of $N$, eq.~(\ref{eq:u-times-vbar}) takes the form
\begin{eqnarray}
  u \otimes \bar{v} &=& {1\over \mathcal{N}}~\slashed{p}_1\frac{1-h\g_5}{2}
         i\ep_{\g p_3 p_4 p_5}\frac{1+h\g_5}{2} \slashed{p_2} \nonumber \\
   &=& {1\over \mathcal{N}} \left(\frac{1}{2}\slashed{p}_1i\ep_{\g p_3 p_4 p_5}\slashed{p}_2
      -\frac{h}{2}\slashed{p}_1\left(\frac{1}{3!}\ep_{\g\g\g \mu}
      \ep^{\mu p_3 p_4 p_5}      \right)
     \slashed{p}_2 \right)\nonumber\\
   &=& {1\over \mathcal{N}}\left( \frac{1}{2}\slashed{p}_1i\ep_{\g p_3 p_4 p_5}\slashed{p}_2
      +\frac{h}{2}\slashed{p}_1\frac{1}{3!}\left( \sum_{i\neq j\neq k = 3,4,5} (-1)^{N_{perm}[i,j,k]}\;
        \slashed{p}_i\slashed{p}_j\slashed{p}_k\right)
     \slashed{p}_2 \right)\,,
     \label{eq:wf-quarks}
\end{eqnarray}
where $N_{perm}[i,j,k]$ is the number of transpositions needed to map an ordering $(i,j,k)$ to the default ordering $(i,j,k)=(3,4,5)$.

As mentioned, in the above equation we have used the relations $\{\g^{\mu},\g_5\} = 0$ and
$\g_5\g^\sigma = \frac{-i}{3!}\ep^{\mu\nu\rho\sigma}\g_\mu \g_\nu\g_\rho$, and
all manipulations in eq.~(\ref{eq:wf-quarks}) have been done in $4$ dimensions.
The normalization factor $\mathcal{N}$ needs to be evaluated only after IR
renormalization, in the context of the finite remainder $\mathcal{F}$. 
For this reason the calculation of $\mathcal{N}$ is straightforward 
and can also be performed in $4$ dimensions.

To summarize the results of this section, the amplitude for a specific helicity configuration eq.~(\ref{eq:amplitude}) is obtained by evaluating the photon polarization vectors according to eq.~(\ref{eq:wf-vector}) and the product of the two quark spinors according to eq.~(\ref{eq:wf-quarks}). The resulting expressions are contracted with the tensor $\Gamma_{\mu\nu\rho}$ which is constructed diagrammatically and whose expression is independent of helicities. In practice only the two helicity configurations shown in eq.~(\ref{eq: helicities}) need to be computed. The generation of the Feynman diagrams is performed with the help of a private software and the calculations of spinor traces and color factors are performed with the help of the program {\tt FORM} \cite{Ruijl:2017dtg}.

\subsection{Reduction to pentagon functions}\label{sec:reconstruction}

In this work we consider four independent amplitude structures. They correspond to the two color factors in eq.~(\ref{eq:color-structures}) and the two helicities in eq.~(\ref{eq: helicities}). We would like to express these structures in terms of transcendental functions and transcendental constants with rational coefficients. Since we assume that the set of functions forms a basis, the evaluation of an amplitude structure is equivalent to the evaluation of all its rational coefficients.
To this end, we have built an automated framework that uses finite-field methods to numerically 
evaluate, interpolate and then reconstruct the exact analytical expressions for these coefficients.
Throughout this section, the term \emph{numerical} should be understood to refer to finite-field numerics.
We will now provide a detailed account of our framework.

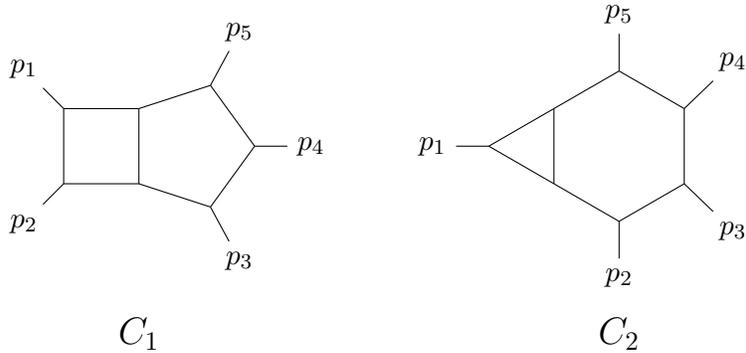
\begin{figure}[]
\centering

\begin{tikzpicture}[scale=0.5]

\node (title) at (0,-5) {\Large $C_1$};
\node (dummy) at (0,-5.5) { };

\node (dummyleft) at (-6,0) { };
\node (dummyright) at (6,0) { };

\node (p1) at  (-3.06066017178,2.06066017178) {$p_1$};
\node (p2) at  (-3.06066017178,-2.06066017178) {$p_2$};
\node (p3) at  (2.67465551853,-3.04461876319) {$p_3$};
\node (p4) at  (4.57768353718,0) {$p_4$};
\node (p5) at  (2.67465551853,3.04461876319) {$p_5$};

\draw (p1) to (-2,1);
\draw (-2,1) to (0,1);
\draw (0,1) to (1.90211303259,1.61803398875);
\draw (1.90211303259,1.61803398875) to (p5);
\draw (1.90211303259,1.61803398875) to (3.07768353718,0);

\draw (p2) to (-2,-1);
\draw (-2,-1) to (0,-1);
\draw (0,-1) to (1.90211303259,-1.61803398875);
\draw (1.90211303259,-1.61803398875) to (p3);
\draw (1.90211303259,-1.61803398875) to (3.07768353718,0);

\draw (-2,1) to (-2,-1);
\draw (0,1) to (0,-1);

\draw (3.07768353718,0) to (p4);
\end{tikzpicture}
\begin{tikzpicture}[scale=0.5]

\node (title) at (0,-5) {\Large $C_2$};
\node (dummy) at (0,-5.5) { };

\node (dummyleft) at (-6,0) { };
\node (dummyright) at (6,0) { };
\node (p1) at  (-4.96410161514,0) {$p_1$};
\node (p2) at  (0,-3.5) {$p_2$};
\node (p3) at  (3.03108891325,-2.25) {$p_3$};
\node (p4) at  (3.03108891325,2.25) {$p_4$};
\node (p5) at  (0,3.5) {$p_5$};

\draw (p1) to (-3.46410161514,0);
\draw (p2) to (0,-2);
\draw (p3) to (1.73205080757,-1);
\draw (p4) to (1.73205080757,1);
\draw (p5) to (0,2);

\draw (0,-2) to (1.73205080757,-1);
\draw (1.73205080757,-1) to (1.73205080757,1);
\draw (1.73205080757,1) to (0,2);
\draw (0,2) to (-1.73205080757,1);
\draw (-1.73205080757,-1) to (-1.73205080757,1);
\draw (-1.73205080757,-1) to (0,-2);

\draw (-1.73205080757,-1) to (-3.46410161514,0);
\draw (-1.73205080757,1) to (-3.46410161514,0);
\end{tikzpicture}
\caption{The integral topologies $C_1$ and $C_2$.
}\label{fig:topologies}

\end{figure}

Any bare scalar 2-loop amplitude, or amplitude structure, $\mathcal{M}$ is expressed as a linear combination of 2-loop scalar integrals $I_b$:
\begin{equation}\label{eq:ampl_to_I}
\mathcal{M} = \sum_b A_{b} I_b.
\end{equation}
The coefficients $A_b$ are rational functions of the kinematic invariants $s_{ij}$, defined in sec.~\ref{sec:notation}, and polynomials in $\ep$. As explained in sec.~\ref{sec:projection} the coefficients $A_b$ are also linear functions of the parity-odd kinematic variable $\tr5$. The planar two-loop integrals $I_b$ appearing in the amplitude of interest in this paper are defined~\cite{Chawdhry:2018awn} as follows:
\begin{equation}\label{eq:C_integrals}
I_b = \int \frac{d^dk_1 d^dk_2 }{\Pi_1^{n_1} \Pi_2^{n_2} \dots \Pi_{11}^{n_{11}}},
\end{equation}
where the propagators $\Pi_i$ are defined in table~\ref{tab:propagators} and $n_i \equiv n_i(b) \in \mathbb{Z}$.
\begin{table}[]
\centering
\caption{The propagators $\{\Pi_i\}$ defining the 2-loop 5-point planar integrals $I_b$}
\label{tab:propagators}
\begin{tabular}{ ccc } 
\toprule
	\hphantom{xxxxxxx} $i$ \hphantom{xxxxxxx}	&	$\Pi_i$		\\
\midrule
1	&	$k_1^2$	\\
2	&	$k_2^2$	\\
3	&	$(k_1+p_1+p_2)^2$	\\
4	&	$(k_1-k_2)^2$	\\
5	&	$(k_2+p_1)^2$	\\
6	&	$(k_2+p_1+p_2)^2$	\\
7	&	$(k_2-p_3)^2$	\\
8	&	$(k_1+p_1+p_2-p_3)^2$	\\
9	&	$(k_1+p_1+p_2-p_3-p_4)^2$	\\
10	&	$(k_2-p_3-p_4)^2$	\\
11	&	$(k_1+p_1)^2		$	\\
\bottomrule
\end{tabular}
\end{table}
These 11 propagators form a complete basis of bilinears through which any scalar numerator structure in an integrand can be expressed.
A maximum of 8 propagators appear in the denominator of any given integrand; for the three remaining propagators, denoted \emph{spurious} propagators, the corresponding indices $n_i$ satisfy $n_i \leq 0$.
We can accordingly classify the integrals into two topologies, which are represented pictorially in fig.~\ref{fig:topologies}.
In the $C_1$ topology, propagators 7, 10, and 11 are spurious. 
In the $C_2$ topology, propagators 6, 7, and 10 are spurious.

Following the standard Integration-By-Parts (IBP) approach, we map the integrals $I_b$ onto a basis of master integrals, $M_c$
\begin{equation}\label{eq:I_to_C1}
I_b = \sum_c B_{bc} M_c\,.
\end{equation}

The coefficients $B_{bc}$ are rational functions of $s_{ij}$ and $\ep$. All coefficients required in this calculation are known analytically from ref.~\cite{Chawdhry:2018awn}. The IBP identities map all required $C_1$ integrals onto a basis of 61 master integrals
\footnote{For the purposes of the IBP-solving approach in Ref.~\cite{Chawdhry:2018awn} there are 62 master integrals in the $C_1$ topology, but two of these are related to each other by a discrete symmetry and can be set equal in the present context.}, 
and all required $C_2$ integrals onto a basis of 28 master integrals. The 28 $C_2$ master integrals can themselves be identified with $C_1$ integrals, in some cases by permuting the external momenta $\{p_j\}$.

The next step in our calculation is to map the master integrals $M_c$ onto a basis of transcendental functions and constants. In this work we choose the basis of functions provided recently in ref.~\cite{Chicherin:2020oor}. That reference solves a set of specially designed loop master integrals $M^{(UT)}$ of uniform transcendentality (UT)
\begin{equation}\label{eq:UT_to_F}
M^{(UT)}_d = \sum_{e} D_{de} t_e,
\end{equation}
where the basis $\{t_e\}$ is built out of sums and products of the transcendental functions \Ffns and transcendental constants \tcr and \tci that are defined in ref.~\cite{Chicherin:2020oor} as well as the unity. The coefficients $D_{de}$ are rational functions of the kinematic invariants $s_{ij}$ and linear functions of $\tr5$. They also depend on $\ep$, however, they are only known as a series expansion of sufficient depth
\begin{equation}\label{eq:UT_to_F_expansion}
D_{de} = \sum_{n=-4}^0 D_{de}^{(n)} \ep^n + \mathcal{O}\left(\ep\right)\,.
\end{equation}

In order to map our set of master integrals $M_c$ onto the function basis $\{t_e\}$ we first use the analytic form of the IBP solutions from ref.~\cite{Chawdhry:2018awn} to express the set $M^{(UT)}$ in terms of the set $M_c$, then we invert it:
\begin{equation}\label{eq:C1_to_UT}
M_c = \sum_d B ^{(UT)}_{cd} M^{(UT)}_d .
\end{equation}
Combining equations~\eqref{eq:C1_to_UT} and~\eqref{eq:UT_to_F}, we obtain an analytical expression for each $C_1$ master integral $M_c$ in terms of the function basis $\{t_e\}$
\begin{equation}\label{eq:C1_to_F}
M_c = \sum_{e} E_{ce} t_e\,,
\end{equation}
where
\begin{equation}
E_{ce} = \sum_d B^{(UT)}_{cd} D_{de}\,.
\end{equation}

We would like to combine the above results in order to express the amplitude $\mathcal{M}$ in terms of the functions $t_e$:
\begin{equation}\label{eq:ampl_to_F}
\mathcal{M} = \sum_e G_{e} t_e\,,
\end{equation}
with coefficients
\begin{equation}\label{eq:ampl_soln}
G_{e} = \sum_b \sum_c A_{b} B_{bc} E_{ce}\,.
\end{equation}

In practice, we only seek the Laurent expansion $G_{e}^{(n)}$ of the coefficients $G_{e}$
\begin{equation}
G_{e} = \sum_{n=-4}^0 G_{e}^{(n)} \ep^n + \mathcal{O}\left(\ep\right)\,,
\label{eq:G-expansion}
\end{equation}
where
\begin{equation}\label{eq:ampl_soln_expanded}
G_{e}^{(n)} = \sum_{n_1+n_2+n_3 = n} \sum_b \sum_c A_{b}^{(n_1)} B_{bc}^{(n_2)} E_{ce}^{(n_3)}\,.
\end{equation}

The coefficient $A_{b}^{(n)}$ appearing in eq.~(\ref{eq:ampl_soln_expanded}) are defined as
\begin{equation}
A_{b} = \sum_{n=n_{\rm min}}^{n_{\rm max}} A_{b}^{(n)} \ep^n \,.
\label{eq:A-expansion}
\end{equation}
The coefficients $B_{bc}^{(n)}$ and $E_{ce}^{(n)}$ appearing in eq.~(\ref{eq:ampl_soln_expanded}) are defined similarly to eq.~(\ref{eq:A-expansion}) but with respect to the functions $B_{bc}$ and $E_{ce}$. For each one of the functions $A_{b}, B_{bc}$ and $E_{ce}$, the powers $n_{\rm min}$ and $n_{\rm max}$ appearing in eq.~(\ref{eq:A-expansion}) are chosen in such a way that the range for the index $n$ in eq.~(\ref{eq:G-expansion}) is satisfied. 

The coefficients $G_{e}^{(n)}$ can be split into two parts: one which is proportional to $\tr5$ and one which is independent of it. Both parts are rational functions of the invariants $s_{ij}$. The reason only the first power of the parity-odd variable $\tr5$ appears in the final result is that $(\tr5)^2$ is itself a polynomial of $s_{ij}$, see eq.~(\ref{eq:tr5^2}). 

The calculation of the coefficients $G_{e}^{(n)}$ is based on eq.~(\ref{eq:ampl_soln_expanded}) and proceeds as follows. We first note that the coefficients $A_{b}^{(n_1)},  B_{bc}^{(n_2)}$ and $E_{ce}^{(n_3)}$ appearing in that equation are all known in analytic form. In principle one can multiply and add them, as required, to derive the analytic expressions of the coefficients $G_{e}^{(n)}$. The problem with this strategy is that the size of the rational expressions that need to be combined becomes huge which hampers their subsequent simplification. Such a strategy was followed by us in ref.~\cite{Chawdhry:2019bji} for the evaluation of the squared amplitude for $q\bar q\to \3g$. We refer the reader to that reference for more details about the subtleties of such an approach. 

In this work we follow an alternative strategy for the evaluation of the coefficients $G_{e}^{(n)}$. The idea is to numerically evaluate the functions $A_{b}^{(n_1)},  B_{bc}^{(n_2)}$ and $E_{ce}^{(n_3)}$, then multiply and add them as appropriate, in order to obtain a numerical value for $G_{e}^{(n)}$ in a given kinematic point. The finite-field evaluations of $G_{e}^{(n)}$ are then passed to the {\tt FireFly} library~\cite{Klappert:2020aqs} in order to reconstruct the exact analytical expressions for the coefficients $G_{e}^{(n)}$.

We have created an automated framework which is designed to calculate simultaneously multiple amplitudes (that have the same kinematics) following the finite-field evaluation approach just described. The reason it may be advantageous to compute several amplitudes at the same time is efficiency, noting that only the coefficients $A_{b}^{(n_1)}$ depend on the amplitude while the coefficients $B_{bc}^{(n_2)}$ and $E_{ce}^{(n_3)}$ are process independent. 

To fully specify our approach we need to describe one more feature which has to do with momentum crossings. The coefficients $A_{b}^{(n_1)}$ do not require any further crossing since all crossings are already included at the diagrammatic level. Similarly, the functions $\{t_e\}$ are already defined in such a way that all possible crossings have been already implemented in their definition \cite{Chicherin:2020oor}. This is one significant difference with respect to the functional basis constructed in ref.~\cite{Gehrmann:2018yef} and used by us in ref.~\cite{Chawdhry:2019bji}. The basis of ref.~\cite{Gehrmann:2018yef} does not include momentum crossings and if they are required the user needs to implement those. As a result of such crossings one generally arrives at a functional basis which is non-minimal. To reduce the extended set of functions to a minimal set, functional identities need to be derived and applied; see ref.~\cite{Chawdhry:2019bji} for more details on this point.

Where momentum crossings still need to be applied is the part of eq.~(\ref{eq:ampl_soln_expanded}) that involves IBPs. Specifically, this affects the coefficients $B_{bc}^{(n_2)}$ as well as the coefficients $E_{ce}^{(n_3)}$ through their constituent coefficients $B ^{(UT)}_{cd}$, see eq.~(\ref{eq:C1_to_UT}). The reason additional crossings are needed in the parts where IBPs are involved is that, in general, an integral can appear in an amplitude with any of the $5!$ permutations of the external legs while the analytical IBP solutions are only needed -- and therefore only provided -- for the `standard', topology-defining permutation eq.~\eqref{eq:C_integrals}, see also table~\ref{tab:propagators} and fig.~\ref{fig:topologies}. In principle any crossing of the IBP solution can be obtained analytically by rearranging the solution in terms of the crossed kinematic invariants. Such a strategy would be impractical in our numerical approach due to the significant size of the IBP solutions and the very large number of crossings. 

The way we deal with crossings in our practical implementation is as follows. The master integral reductions $E_{ce}^{(n)}$ are derived analytically for all required momentum crossings. The coefficients $B_{bc}^{(n_2)}$ are permuted numerically by first applying the corresponding permutation to the numerical values of the kinematic invariants $s_{ij}$ and then evaluating the `standard' IBP solutions at the resulting numerical point. Since the coefficients $B_{bc}^{(n_2)}$ are universal, at each finite-field point they only need to be evaluated once per required momentum crossing, even if an integral $I_b$ appears in several amplitudes. Due to the large number of integrals and crossings, it is impractical to store the full set of numerical values $B_{bc}^{(n_2)}$, even for just a single finite-field point per computing thread
\footnote{We use a multi-core computing cluster. Each computing thread is assigned independent finite-field points and evaluates all the amplitudes at the assigned points, one point at at time.}.
Instead, after evaluating the coefficients $B_{bc}^{(n_2)}$, we immediately multiply by $A_{b}^{(n_1)}$ for all amplitudes $\mathcal{M}$ and we only store running totals of $\sum_b A_{b}^{(n_1)} B_{bc}^{(n_2)}$.

\section{Results}\label{sec:results}

Following the approach described in the previous section, in this work we calculate in analytical form the two helicities eq.~(\ref{eq: helicities}) of the finite remainder for the process $q\bar q\to \3g$ in the 't Hooft-Veltman scheme. In this work we have not included any non-planar contributions i.e. we work in the approximation eq.~(\ref{eq:F2-decomposition-LC}). Factoring out the phase-dependent part of the leading order amplitude we write the reconstructed finite remainder for the helicity $\bar{h}_+$ defined in eq.~(\ref{eq: helicities}) as:
\begin{equation}
\mathcal{F}^{\bar{h}_+} = \mathcal{F}^{\bar{h}_+ (0)}\biggl(1+\frac{\alpha_s}{4\pi} C_F\mathcal{R}^{\bar{h}_+ (1)}
 +\biggl(\frac{\alpha_s}{4\pi}\biggr)^2 \biggl(N_c^2\mathcal{R}^{\bar{h}_+ (N_c^2)}+C_F n_f\mathcal{R}^{\bar{h}_+ (C_F n_f)} \biggr) + {\cal O}(\alpha_s^3) \biggr)\,.
 \label{eq:result+}
\end{equation}
For the helicity $\bar{h}_-$ the tree-level amplitude vanishes and we write
\begin{equation}
\mathcal{F}^{\bar{h}_-} = \tilde{\mathcal{F}}^{\bar{h}_-}\biggl(\frac{\alpha_s}{4\pi} C_F\mathcal{R}^{\bar{h}_- (1)}
 +\biggl(\frac{\alpha_s}{4\pi}\biggr)^2 \biggl(N_c^2\mathcal{R}^{\bar{h}_- (N_c^2)}+C_F n_f\mathcal{R}^{\bar{h}_- (C_F n_f)} \biggr) + {\cal O}(\alpha_s^3) \biggr)\,.
  \label{eq:result-}
\end{equation}
The functions $\mathcal{R}^{\bar{h} (i)}, i=(1, N_c^2, C_F n_f)$, have the following structure
\begin{equation}
 \mathcal{R}^{\bar{h} (i)} = \sum_e r^{\bar{h} (i)}_e t_e\,.
 \label{eq:R}
\end{equation}
They are independent of the specific phase choices made in the construction of the external wave-functions. The coefficients $r^{\bar{h} (i)}_e$ appearing in eq.~(\ref{eq:R}) are rational functions of the parity-even invariants $s_{ij}$ and are linear functions of $\tr5$. The explicit expressions for the functions $\mathcal{R}^{\bar{h} (i)}$, as well as for the tree-level amplitude $\mathcal{F}^{\bar{h}_+ (0)}$ and for the phase-dependent factor $\tilde{\mathcal{F}}^{\bar{h}_-}$ can be found in electronic form in the ancillary files accompanying the arXiv submission of this article.

We have checked that the above result agrees numerically with our previous calculation performed in ref.~\cite{Chawdhry:2019bji}. This comparison does not include the terms $\sim n_f$ since those were not computed in ref.~\cite{Chawdhry:2019bji}. Given the two calculations were performed with almost completely independent methods this represents a strong check on the correctness of eqs.~(\ref{eq:result+},\ref{eq:result-}). Since one of the main objectives of the present work is to document the calculation of the two-loop amplitude used in ref.~\cite{Chawdhry:2019bji}, we will next explain in some detail how the two calculational approaches differ from each other. 

Ref.~\cite{Chawdhry:2019bji} computed directly the squared amplitude, while in this work we compute the helicity amplitudes and the squared amplitude is obtained by squaring and crossing them numerically. Ref.~\cite{Chawdhry:2019bji} used the polygon functional basis of ref.~\cite{Gehrmann:2018yef} while in the present work we use the basis of ref.~\cite{Chicherin:2020oor}. The main difference between the two functional bases was explained in the previous section. Our calculation in ref.~\cite{Chawdhry:2019bji} was fully analytic while the current calculation uses finite-field numeric evaluation and reconstruction techniques to obtain the rational coefficients. Besides the differences already mentioned, the two results differ significantly as far as evaluation times are concerned. The main reason for this is the difference in evaluation times and numerical precision between the two functional bases. The larger size of the result in ref.~\cite{Chawdhry:2019bji} is immaterial given the time needed for the evaluation of the most complicated functions in that basis. 

We have also checked that our results eqs.~(\ref{eq:result+},\ref{eq:result-}) agree with the results in ref.~\cite{Abreu:2020cwb}. Since we have computed a different helicity combination relative to the helicities published in ref.~\cite{Abreu:2020cwb}, a direct analytic comparison between the two is complicated by the fact crossing of external legs is required. We have analytically checked most structures which are simple enough to cross, while the complete expressions (for each helicity and color factor) have been compared numerically with high numerical precision (30 digits) and full agreement between the two calculations has been found in all cases. There are certain differences in the way the present calculation and the one in ref.~\cite{Abreu:2020cwb} were performed that make the agreement between the two calculations highly nontrivial. First, the generation of diagrams is based on different approaches. Furthermore, in this work we use the analytic solutions of the IBP equations from ref.~\cite{Chawdhry:2018awn} while ref.~\cite{Abreu:2020cwb} solves the IBP equations numerically for each point in which the amplitude is evaluated and being reconstructed. Most importantly, in this work we use an alternative approach to the projection of helicity amplitudes which is very different from the one employed in ref.~\cite{Abreu:2020cwb}.

\section{Conclusion}\label{sec:conclusion}

In this work we calculate the planar contributions to the two-loop helicity amplitude for the process $q\bar q\to \3g$. The result is presented in fully analytic form and is available for download in electronic form with the arXiv submission of this paper. This result, written in an alternative form, was used in the first NNLO calculation of a $2\to 3$ LHC process \cite{Chawdhry:2019bji}.

The helicity amplitudes are expressed in terms of the functional basis of ref.~\cite{Chicherin:2020oor} which allows fast and efficient numerical evaluation of the amplitudes. These functions' speed of evaluation is sufficiently high to allow the direct use of this amplitude in the calculation of NNLO cross-sections without the need for intermediate interpolation. 

In our calculation we have utilized Chen's recently proposed approach \cite{Chen:2019wyb} for the efficient projection of helicity amplitudes in multileg/multiloop processes. We have found the approach very easy to implement and use especially since it does not involve investigations of tensor bases that grow with the number of loops. 

We have found complete agreement between our calculation and the recent independent calculation of the same amplitude in ref.~\cite{Abreu:2020cwb}. Such an agreement represents a highly nontrivial check on both calculations given they use very different computational approaches. 

Our calculation is derived from an automated framework we have created for the calculation of generic two-loop massless 5-point gauge theory amplitudes. We hope it will prove suitable for many other potential applications.

\begin{acknowledgments}
We acknowledge helpful discussions with Long Chen about the helicity projection method of ref.~\cite{Chen:2019wyb} and with Fabian Lange and Jonas Klappert on the use of the library {\tt FireFly} \cite{Klappert:2020aqs}.
The work of M.C. was supported by the Deutsche Forschungsgemeinschaft under grant 396021762 - TRR 257. The research of H.C., A.M. and R.P. has received funding from the European Research Council (ERC) under the European Union's Horizon 2020 Research and Innovation Programme (grant agreement no. 683211). A.M. was also supported by the UK STFC grants ST/L002760/1 and ST/K004883/1. The research of H.C. has also received funding under the ERC grant agreement 804394. A.M. acknowledges the use of the DiRAC Cumulus HPC facility under Grant No. PPSP226.
\end{acknowledgments}

\appendix
\section{Renormalization constants}\label{sec:app-renorm}

The UV renormalisation constant of the quark wave function
through order ${\cal O}(\alpha_s^2)$ reads
\begin{equation}
 Z_q = 1 + \biggl(\frac{\alpha_s}{4\pi}\biggr)^2 \left(C_F T_F \left(-\frac{5}{6}
        + \frac{1}{\ep}\right)\right)\;,
\end{equation}
while the renormalization constant $Z_{\alpha_s} = Z_g^2$ up to power $\alpha_s$
(higher powers are not required since the tree-level amplitude for $pp\to\gamma\gamma\gamma$ has a zero power of $\alpha_s$)
is given by
\begin{equation}
  Z_g = 1 + {1\over \ep} \biggl(\frac{\alpha_s}{4\pi}\biggr) \frac{4(n_l+1) T_F-11 C_A}{6} \,.
\end{equation}
The contribution from heavy flavours is not considered in this work, therefore we have $Z_q = 1$ and the term $(n_l+1)$ term in $Z_g$ reads $n_l$.

For the IR renormalization of the amplitude for the process $q\bar{q} \to \g\g\g$ the color-space matrix $\mathbf{Z}$ is needed. Through order ${\cal O}(\alpha_s^2)$ it is given  by
\begin{equation}
\begin{split}
 \mathbf{Z} &= 1 + \biggl(\frac{\alpha_s}{4\pi}\biggr)\left(
  -\frac{2 C_F}{\ep^2}-
   \frac{ 2 C_F l_\mu+3 C_F}{\ep}
   \right)\\
  &+\biggl(\frac{\alpha_s}{4\pi}\biggr)^2 \biggl(
  \frac{2 C_F^2}{\ep^4}+
  \frac{C_F (11 C_A+8 C_F l_\mu+12
        C_F-4 n_l T_F)}{2 \ep^3}\\
  &+\frac{C_F \left(6 l_\mu (11 C_A+18 C_F-4 n_l T_F)+2 C_A
        (9 \zeta_2+16)+36 C_F l_\mu^2+81 C_F-16 n_l
        T_F\right)}{18 \ep^2}\\
 &+\frac{C_F (12 l_\mu (C_A (18
\zeta_2-67)+20 n_l T_F)+C_A (1404 \zeta_3-594 \zeta_2-961)-1296 C_F \zeta_3}{108 \ep}\\
   &+\frac{648 C_F \zeta_2-81 C_F+216 n_l T_F \zeta_2+260 n_l T_F)}{108 \ep}
 \biggr)\;,
\end{split}
\end{equation}
with the abbreviation $l_\mu = \log (-\mu^2/s_{12})$.

\end{document}